\documentclass[twocolumn]{emulateapj}

\usepackage{natbib}
\usepackage{epsf}

\def\lsim{\mathrel{\rlap{\lower4pt\hbox{\hskip1pt$\sim$}}
    \raise1pt\hbox{$<$}}}                

\shorttitle{The SFR history in the FDF and GOODS-South fields}
\shortauthors{Gabasch et al.}

\begin{document}


\title{The star formation rate history in the  FORS Deep and GOODS
  South Fields
\altaffilmark{1}}
\altaffiltext{1}{Based on observations collected at the European Southern
  Observatory, Chile (ESO programmes 
  63.O-0005, 64.O-0149, 64.O-0158, 64.O-0229, 64.P-0150, 65.O-0048,
  65.O-0049, 66.A-0547, 68.A-0013, and 69.A-0014)}

\author{A. Gabasch,  M. Salvato, R.P. Saglia, R. Bender,  U. Hopp,  S. Seitz, G. Feulner,  M. Pannella}
\affil{Max-Planck-Institut f\"ur extraterrestrische Physik, Giessenbachstr., Postfach 1312, D-85741 Garching, Germany}
\affil{Universit\"atssternwarte M\"unchen, Scheinerstr. 1, D-81673 M\"unchen, Germany}
\and
\author{N. Drory}
\affil{University of Texas at Austin, Austin, Texas 78712}
\and
\author{M. Schirmer, T. Erben}
\affil{Institut f\"ur Astrophysik und extraterrestrische Forschung, Universit\"at Bonn, Auf dem H\"ugel 171, D-52121 Bonn, Germany}




\begin{abstract}
  We measure the star formation rate (SFR) as a function of redshift
  $z$ up to $z\approx 4.5$, based on B, I and (I+B) selected galaxy
  catalogues from the FORS Deep Field (FDF) and the K-selected
  catalogue from the GOODS-South field.  Distances are computed from
  spectroscopically calibrated photometric redshifts accurate to
  $\Delta z / (z_{spec}+1) \le 0.03$ for the FDF and $\le 0.056$ for
  the GOODS-South field. The SFRs are derived from the luminosities at
  1500~\AA. We find that the total SFR estimates derived from B, I and
  I+B catalogues agree very well ($\lsim 0.1$ dex) while the SFR from
  the K catalogue is lower by $\approx 0.2$ dex. We show that the
  latter is solely due to the lower star-forming activity of
  K-selected intermediate and low luminosity ($L<L_\ast$) galaxies.
  The SFR of bright ($L>L_\ast$) galaxies is independent of the
  selection band, {\it i.e.}  the same for B, I, (I+B), and K-selected
  galaxy samples.  At all redshifts, luminous galaxies ($L>L_\ast$)
  contribute only $\sim \frac{1}{3}$ to the total SFR.  There is no
  evidence for significant cosmic variance between the SFRs in the FDF
  and GOODs-South field, $\lsim 0.1$ dex, consistent with theoretical
  expectations.  The SFRs derived here are in excellent agreement with
  previous measurements provided we assume the same faint-end slope of
  the luminosity function as previous works ($\alpha\sim -1.6$).
  However, our deep FDF data indicate a shallower slope of
  $\alpha=-1.07$, implying a SFR lower by $\approx 0.3$ dex.  We find
  the SFR to be roughly constant up to $z\approx 4$ and then to
  decline slowly beyond, if dust extinctions are assumed to be
  constant with redshift.
\end{abstract}



\keywords{galaxies: high-redshift --- galaxies: formation --- galaxies: evolution}


\section{Introduction}
\label{sec:sfr:introduction}

The determination of the star formation rate (SFR) history of the
universe is one of the most interesting results extracted from the
deep photometric and spectroscopic surveys of the last decade. A large
number of measurements have been collected, at low \citep[the
Canada-France redshift survey at $z<1$,][]{lilly:1}, and high redshift
from the Hubble Deep Field North \citep{madau:96}, the large samples
of U and B drop-out galaxies \citep{steidel:1}, up to the most recent
determinations based on I-dropouts at redshift $\approx 6$ from the
GOODS, UDF and UDF-Parallel ACS fields \citep{giavalisco:2,bunker:1,
  bouwens:2004}. These studies show that the SFR (uncorrected for
dust) increases from $z=0$ to $z=1$, stays approximately constant in
the redshift range $1-4$, and starts to decline at larger redshifts.
In all the cases quoted above the determination is based on the
estimate of the total UV galaxy luminosity density, that for a given
IMF is proportional to the instantaneous SFR
\citep{madau:96,mad_poz_dick1}.  As discussed by many authors (e.g.
\citealt{hopkins:1}) this approach is affected by the uncertainties of
dust correction, but roughly agrees with other estimators at low to
intermediate redshifts ($z\le1$). Theoretical models of galaxy
formation and evolution can be tested against the measured SFR history
\citep{somerville,hernquist:1}.

So far, all determinations of the SFR history have suffered from some
major limitations. High redshift samples have been small in number due
to the limited field of view of deep pencil-beam surveys, resulting in
large Poissonian fluctuations and large field-to-field variations 
(cosmic variance). The faint-end of the luminosity function (LF)
is thusfar only poorly constrained at high redshifts, implying large
completeness correction factors.  Finally, the technique used to
generate the high-redshift galaxy catalogues (drop-out selection,
optical magnitude limited survey) might have introduced biases by
selecting only specific types of galaxies and possibly missing
relevant fractions of UV light \citep{ilbert:1}.

Here we try to minimize these uncertainties and determine the SFR
history of the universe with improved accuracy up to $z\approx 4.5$.
Our sample of high redshift galaxies is based on two deep fields, the
(I and B selected) FORS Deep Field \citep[FDF, ][]{fdf_data}, and the
(K-selected) GOODS-South field \citep{giavalisco:1}.  Both cover a
relatively large sky area, reducing the problem of cosmic variance.
Both are deep enough to allow the detection of several \mbox{$\times$
  $10^3$} galaxies, thus minimizing the effect of shot noise.

Accurate photometric redshifts ($\Delta z / (z_{spec}+1) \le 0.03$ for
the FDF and $\le 0.056$ for the GOODS-South field) with only $\approx
1$ \% catastrophic failures allow us to measure the UV
  luminosity function down to fainter limits than spectroscopic
  samples.  A detailed comparison of the UV luminosity
  functions of the FDF with the LF derived in large surveys was
  presented in \citet{gabasch:1} and shows good agreement in the
  overlapping magnitude range at all redshifts.  Finally and
  most importantly, the two fields provide us with B-band, I-band and
K-band selected catalogues, making it possible to assess the
dependence of the SFR on the detection band and galaxy colors and the
associated selection biases.

The Letter is organized as follows.  In \S \ref{sec_photometry} we
discuss the photometry and the photometric redshifts of the two
fields, in \S \ref{sec_sfr} we present our results on the SFR history,
and in \S \ref{sec_conclusions} we draw our conclusions.  Throughout
the paper we use AB magnitudes and adopt a concordance cosmology with
\mbox{$\Omega_M=0.3$}, \mbox{$\Omega_\Lambda=0.7$}, and \mbox{$H_0=70
  \, \mathrm{km} \, \mathrm{s}^{-1} \, \mathrm{Mpc}^{-1}$}.

\section{Data sets}
\label{sec_photometry}

The present results are based on photometric catalogues derived for
the FDF \citep[][ UBgRI,\mbox{834~nm},zJKs bands]{fdf_data,gabasch:1} and the
GOODS-South fields \citep[][ UBVRIJHKs bands]{salvato:1}.  
The two
fields cover approximately the same area (39.81 arcmin$^2$ for FDF and
50 arcmin$^2$ for GOODS); the FDF reaches effective absolute magnitude
limits $\approx 1$ mag deeper than GOODS-South (see below).  We use
the I-band and B-band selected FDF catalogues as derived in
\citet{fdf_data} and \citet{gabasch:1}.  The B and I selected
catalogues list 5488 and 5557 bona-fide galaxies (having excluded the
one known bright quasar in the field) down to $B_{lim}=27.6$ and
$I_{lim}=26.8$, respectively. The I+B catalogue obtained by merging
these two contains 6756 entries. 
Photometric redshifts for the FDF galaxies have an accuracy of
\mbox{$\Delta z / (z_{spec}+1) \le 0.03$} with only $\sim 1$\% catastrophic
outliers \citep{gabasch:1}.

Our K-band selected catalogue for the GOODS-South field is based on
the 8 $2.5\times2.5$ arcmin$^2$ J, H, Ks VLT-ISAAC images publically
available, taken with seeing in the range $0.4''-0.5''$.  The U and I
images are from GOODS/EIS public survey, while B V R are taken from
the Garching-Bonn Deep Survey. Data reduction is described in
\citet{arnouts:1} and \citet{schirmer:1}, respectively.
  The data for the GOODS field were analyzed in a very similar way to
  the data of the FDF.  The objects were detected in the K-band images
  closely following the procedure used for the FDF I and B band
  detection \citep{fdf_data}, using both SExtractor \citep{bertin} and
  the YODA package \citep{drory:3}.  A detailed description of the
  procedure can be found in \citet[][]{salvato:1}.  We detected 3367
  objects in K-band for which we derived magnitudes (fixed aperture
  and total) in all bands.  Number counts match the literature values
  down to $K\approx 25.4$, which is the completeness limit of the
  catalogue, in agreement with the number obtained following
  \citet{snigula:1}.  Note that much deeper ACS based catalogues are
  available \citep{giavalisco:1}, but as we are focusing on
  the K-selection they are not relevant in this context. We computed
  photometric redshifts following \citet[ 2004, in
  preparation]{bender:1} and using the same SED template spectra as
  for the FDF.  The comparison with the spectroscopic redshifts of the
  VIMOS team \citep{lefevre:1} and the FORS2 spectra released at
  http://www.eso.org/science/goods/, shows that the photometric
  redshifts have an accuracy $\Delta z /(z_{spec}+1) \le 0.056$.
  Similar results are obtained when comparing to the COMBO-17
  \citep{wolf:1} data.
  This is nearly a factor of 2 better than \citet{mobasher:1} obtained
  using ground-based plus HST/ACS data.  Stars are identified and
  excluded as in \citet{gabasch:1}, as well as known AGN
  \citep{szokoly:1}, leaving 3297 bona-fide galaxies used in the
  further analysis.

\begin{figure}
\epsscale{1.1}
\plotone{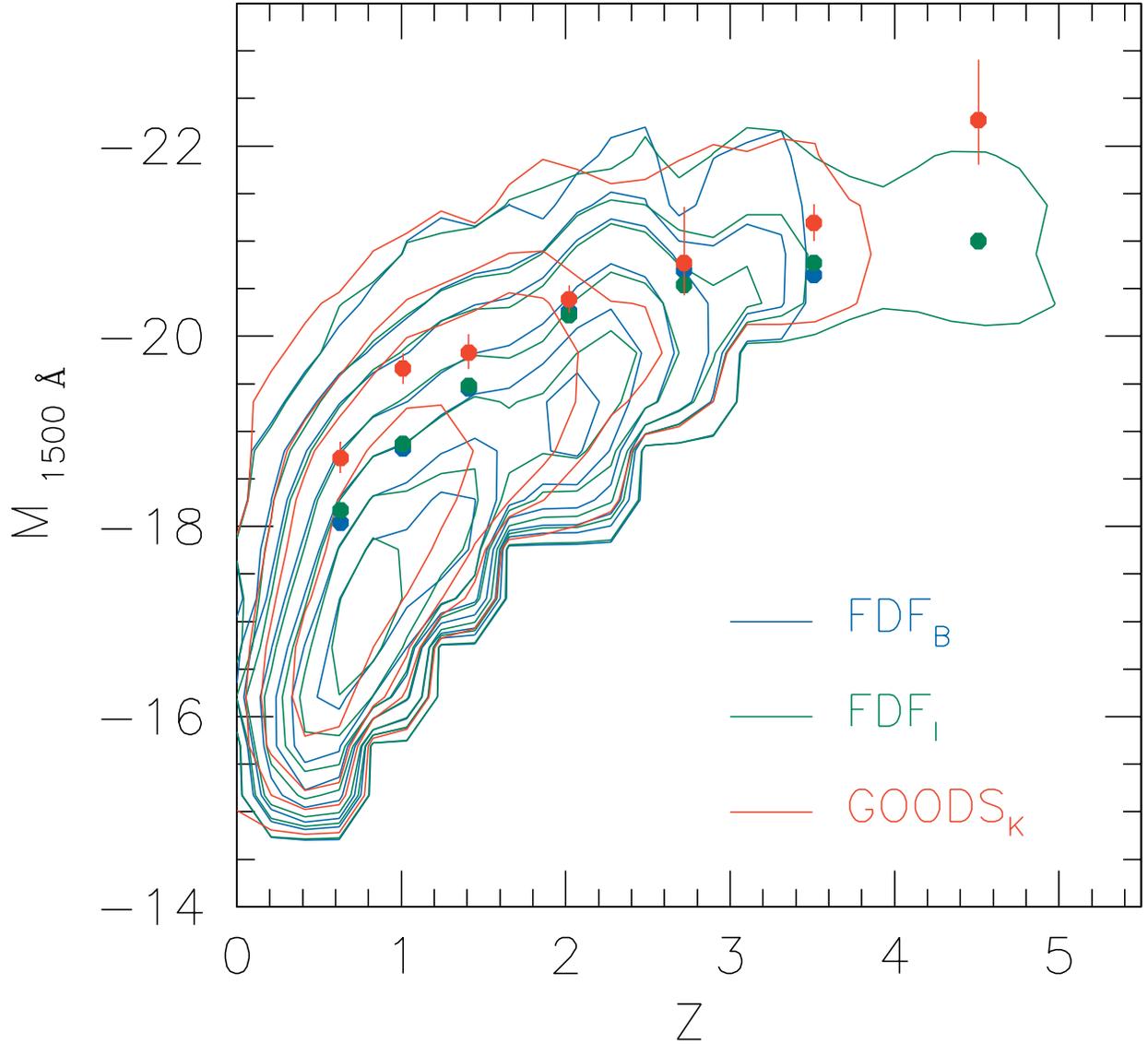}
\caption{The distribution of galaxies in the rest-frame, 1500
\AA\ absolute magnitude vs. redshift space, slightly smoothed with a Gaussian
kernel. The red colors refer to the K-selected galaxies of the
GOODS-South field, the blue and the green colors to B and I selected
galaxies of the FDF. The lowest contour corresponds to
0.75 galaxies/arcmin$^2$/mag per unit redshift bin; the others give
the 2.5, 3.75, 6.25, 8.75, 11.25 and 13.75 galaxies/arcmin$^2$/mag per
unit redshift bin density levels.  
For a  better comparison of the
FDF and GOODS-South samples at the faint-end, we chose 
the completeness limit of the
GOODS-South as 
the magnitude cut-off for all samples. 
The solid circles show the best-fit values
of $M_\ast$, with the errorbars of the K determinations (similar or smaller 
errors are derived in I and B).
\label{fig:sfr:cont_absmag_fdf_goods}}
\end{figure}

  Fig. \ref{fig:sfr:cont_absmag_fdf_goods} shows the distribution of
  galaxies (slightly smoothed with a Gaussian kernel) in the rest
  frame $1500 $\AA\ absolute magnitude $M_{1500}$ vs. redshift
  plane, computed by integrating the best fitting SED over the band
  definition ($1500\pm 100 $\AA).  The contours agree remarkably well
  at the bright-end showing that the number density of bright galaxies
  does not significantly depend on the wavelength at which they were
  selected (for the B-band, this is of course only true up to
  $z\approx 3$). For better comparability of the FDF and GOODS-South
  samples at the faint-end, we chose a consistent magnitude cut-off
  for all samples in Fig. \ref{fig:sfr:cont_absmag_fdf_goods}. This
  magnitude cut-off corresponds to the completeness limit of the
  GOODS-South sample and is about one magnitude brighter than the
  completeness limits of the FDF B and I samples. For the redshift
  bins defined by the limits 0.45, 0.81, 1.21, 1.61, 2.43, 3.01, 4.01,
  5.01, the cut-offs in $M_{1500}$ are at -15, -16, -17, -18, -19,
  -20, -20.
  
  The $M_{1500}$ LFs and the related parameters $M_\ast$, $\Phi_\ast$
  and $\alpha$ of the B, I, and (I+B) selected FDF samples are almost
  identical.  We derived consistent values for $\alpha$
  (\mbox{$-1.07\pm 0.04$}) for all three samples considered here,
  similar to that described in \citet{gabasch:1}. 
  Consistent faint-end slopes ($\alpha = -1.01 \pm 0.08$) were
  obtained using a brighter subset of the data set (i.e. 1 mag
  brighter than the 50\% completeness limit).  Objects that were
  detected in only one band and not in both are all faint and do not
  contribute significantly to SFR determined from the integral over
  the LF.  \citet{gabasch:1} show that the steeper slope of other
  surveys is largely due to shallower limiting magnitudes.  This is
  supported by an analysis of \mbox{$z \sim 6$} dropouts from GOODS
  and the Hubble Ultra Deep Parallel Fields \citep{bouwens:2004} where
  an $\alpha$ of -1.15 was derived.  Compared to LF parameters of the
  optically selected samples, the $M_{1500}$ LF of the K-selected
  sample has slightly brighter values of $M_\ast$, significantly lower
  values in $\Phi_\ast$ and, within the large errors, a similar
  faint-end slope $\alpha$.  Since the slightly shallower K-selected
  sample does not allow us to constrain the faint-end slope to the
  same level as our FDF sample (but is consistent with the faint end
  slope $\alpha=-1.07$ determined for that field), we adopt this value
  for our K-selected sample.

\noindent
We examine the consequences of these findings for the SFR in the next
section.

\section{The Star Formation Rate}
\label{sec_sfr}

\begin{figure*}
\epsscale{1.}
\plotone{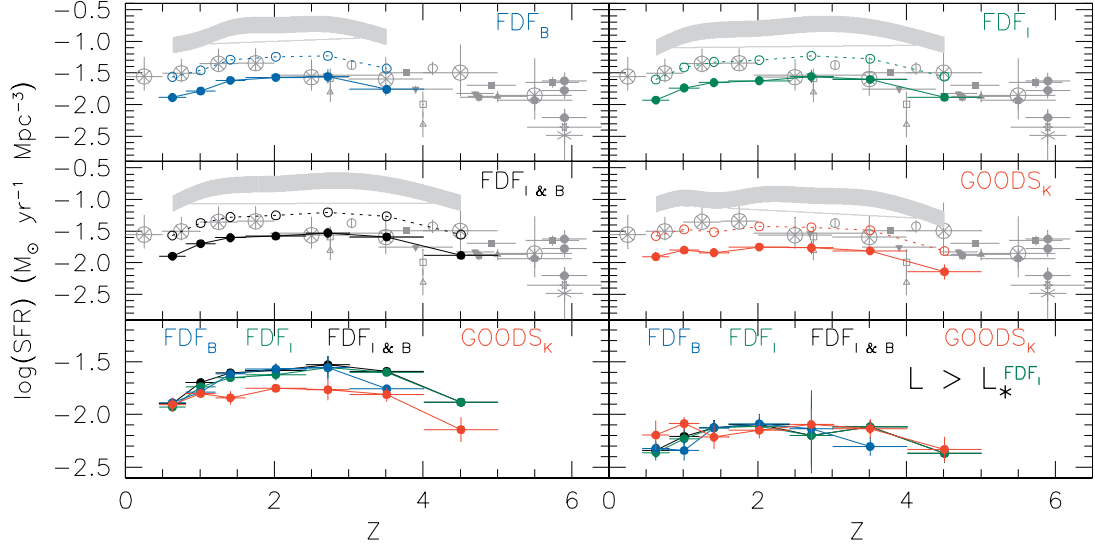}
\caption{The four plots at the top show the SFR as a function of
redshift derived from the 1500 \AA\ luminosity densities computed
from the B-selected (blue), I (green) and I+B-selected (black) FDF, and
K-selected (red) GOODS-South field. The points are connected by the
thick lines for clarity. These SFRs are based on a faint-end slope of the LF 
of -1.07 as derived from the FDF and GOODS data.  The
dotted lines show the effect of assuming a slope of -1.6. 
The grey-shaded  region shows the effect of dust corrections 
with correction factors between 5 and 9, following \citet{adelberger:1}.  
The grey symbols show the results 
\citep[taken from the table of][]{somerville} 
of \citet[ circled crosses]{plf}, 
\citet[ open circles]{steidel:1}, 
\citet[ open triangles]{madau:96},
\citet[ open squares]{mad_poz_dick1}, and 
\citep[taken from ][]{bunker:1} 
\citet[ filled triangles]{iwata:1}, 
\citet[ filled squares]{giavalisco:2}, 
\citet[ filled circles]{bouwens:2003}, 
\citet[ hexagonal crosses]{bouwens:2004}, \citet[ filled pentagons]{fontana:1}, 
\citet[ open star]{bunker:1}, 
\citet[ inverted filled triangles]{bbi:1}. The plots at
the bottom show the SFRs of the four catalogues together (left) and
the SFRs derived considering the contributions of the galaxies
brighter than $L_\ast^I$ only (right).
\label{fig:sfr:sfr_fdf_goods}}
\end{figure*}

We compute the SFR for all 4 catalogues from the total luminosity
densities $l_{1500}$ in the 1500 \AA\ band.  First, we derive
$l_{1500}$ at a given redshift by summing the completeness corrected
\citep[using a $V/V_{max}$ correction, see][]{gabasch:1} LFs up to the
\mbox{1500 \AA\ } absolute magnitude limits.  Second, we apply a
further correction (to zero galaxy luminosity) ZGL, to take into
account the missing contribution to the luminosity density of the
fainter galaxies.  To this end we use the best-fitting Schechter
function.  For the FDF catalogues the ZGL corrections are only 2-20\%
in size.  The small ZGL correction employed here owes itself to the
faint magnitude limits probed by our deep FDF data set and the
relatively flat slopes ($\alpha\approx -1.07$) of the Schechter
function.  Due to the brighter magnitude limit, the ZGL corrections
for the GOODS catalogue can be as high as 50\%.  Note that if we
follow i.e.  \citet{steidel:1} who find $\alpha=-1.6$ \citep[excluded
at 2$\sigma$ with our fits, see ][]{gabasch:1}, we would get much
larger ZGL corrections for the same $M_\ast$, $\Phi_\ast$ (see the
dotted line in Fig.~\ref{fig:sfr:sfr_fdf_goods} and the discussion
below).

Finally, following \citet{mad_poz_dick1} we derive the SFR by scaling
the UV luminosity densities: ${\rm
  SFR}_{1500}=1.25\times10^{-28}\times l_{1500}$ in units of $M_\odot
yr^{-1}Mpc^{-3}$, where the constant is computed for a Salpeter IMF.
The resulting values of ${\rm SFR}_{1500}$ 
are shown in Fig.  \ref{fig:sfr:sfr_fdf_goods} as a function of
redshift.  Errors are computed from Monte Carlo simulations that take
into account the probability distributions of photometric redshifts
and the Poissonian error \citep{gabasch:1}.  Following
\citet{adelberger:1}, we assume that dust extinction does not evolve
with redshift and is about a factor of $\sim 5-9$ in the rest-frame
UV. A more detailed discussion of the role of dust will be given in a
future paper, like an analysis based on the SFR derived at 2800 \AA.
Thanks to the large area covered and the faint limiting magnitudes
probed, our determination of the SFR is the most precise to date, with
statistical errors less than 0.1 dex for the single redshift bins
spanning the range $0.5< z<5$.

The considerations of \S\ref{sec_photometry} translate in the
following conclusions about the SFR. Out to redshift $z\approx 3$ the
SFRs derived from the I and B selected FDF, or the merged I+B
catalogue, are identical within the errors ($\lsim 0.1$ dex; see plot
at the bottom left of Fig.  \ref{fig:sfr:sfr_fdf_goods}). At larger
redshifts the B-selected SFRs underestimate the true values, since B
drop-outs are not taken into account. The strong evolution in both the
$M_\ast$ and $\phi_\ast$ parameters of the Schechter LF measured as a
function of redshift by \citet{gabasch:1} results in a nearly constant
SFR, because the strong brightening of $M_\ast$ is compensated by the
dramatic decrease of $\phi_\ast$ with $z$.
Comparing the two lower panels of \mbox{Fig.
  \ref{fig:sfr:sfr_fdf_goods}} shows that luminous galaxies
($L>L_\ast$) contribute only a third of the total SFR at all observed
$z$, independent of the selection band.

The K-selected SFRs are similar in shape, but systematically lower by
$\approx 0.2$ dex at $z>1$. This result holds independently of our
completeness correction.  If we consider only the contributions to the
SFR down to the limiting magnitude set by the K-band, we find the same
0.2 dex difference for $1<z\le 3$, and 0.15 dex at $z> 3$.  Fig.
\ref{fig:sfr:cont_absmag_fdf_goods} shows that this result originates
from the lower density of $M_{1500}>-19$ galaxies in the K-selected
catalogue, as intermediate and low luminosity blue galaxies
contributing to the SFR budget are more easily detected in the bluer
bands than in K. In fact, the contributions to the SFR coming from the
galaxies brighter than $L_\ast^I$ are identical within the errors for
the I and K selected catalogues (see Fig.
\ref{fig:sfr:sfr_fdf_goods}, bottom-right panel).  Therefore, cosmic
variance does not play a role, as we also verified by comparing the
B-band number counts in the 2 fields. They agree within 0.1 dex, which
is the expected variation derived by \citet{somerville:2} scaled to
the area of the GOODS-South field.
On the other hand, \citet{gabasch:1} show that the I-band FDF
catalogue might be missing only about 10~\% of the galaxies that would
be detected in a deep K-band selected survey with magnitude limit
$K_{AB}\approx 26$ \citep[like in][]{labbe:1}. The missing galaxies
would be faint and likely not contributing significantly to the SFR
provided their dust extinction is not exceedingly large.  Independent
of the selection band the SFR declines beyond $z \sim 4.5$.  Our
results confirm the conjecture of \citet{kashikawa:1} that the
K-selected UV LFs match the optically selected LFs at high
luminosities.

The comparison with the literature shows that our results are
\mbox{$\sim 0.3$} dex lower, independent of the selection band.
This difference stems from the large completeness
corrections applied by, e.g., \citet{steidel:1}, derived from the
steep slopes fitted to the LF (see \S\ref{sec_photometry}). Our
results scale to the literature values if similar slopes 
are used for the same $M_\ast$ and $\phi_\ast$.  This is
shown by the dotted lines of Fig. \ref{fig:sfr:sfr_fdf_goods}, where
we have assumed a slope of $-1.6$ for our data set while keeping
$M_\ast$ and $\phi_\ast$ the same as in our fit.

The overall agreement between the SFRs derived over a wide wavelength
range (within 0.2 dex),
from the optical B and I to the NIR K, sampling at $z\approx 4$ the
rest-frame UV and B, shows that we are approaching (in the optical)
the complete census of the galaxies contributing to the stellar
production of the universe up to this redshift. Therefore, we can
expect possible biases induced by missing stellar energy distributions
with redshift \citep{ilbert:1} to be small, when deep enough optical
or NIR catalogues are available.  However, we might still not take
into account the possible contribution to the SFR coming from faint,
highly dust-absorbed red star-forming galaxies
\citep{hughes:98,genzel:1} which are likely missing from optically or
near-infrared selected samples.
Nevertheless, it is encouraging to find that recent Spitzer results
(e.g. \citealt{egami:1}) indicate that the majority of the star
formation has already been accounted for using the dust-corrected SFR
derived from optical studies.

\section{Conclusions}
\label{sec_conclusions}

We have measured the SFR of the universe out to $z\approx 4.5$ with
unprecedented accuracy from the FORS Deep Field and the GOODS-South
Field ($90$ arcmin$^2$ in total). Our main conclusions are:\\
$\bullet$ The cosmic variance in the SFR history between the FDF and
GOODS-South field is negligibly small.  The difference between these
fields is $\lsim 0.1$ dex,
consistent with theoretical expectations.\\
$\bullet$ The SFR of galaxies brighter than $L_\ast^I$ is the same
($\lsim 0.1$ dex) in B, I, (I+B) and K selected catalogues.  This
indicates that present optical and NIR surveys are unlikely to have
missed a substantial fraction population of massive star forming
objects (with the possible exception of heavily dust-enshrouded
starbursts).\\
$\bullet$ The total SFR integrated over all galaxy luminosities is the
same in the B, I, and (I+B) selected catalogues and is lower in the
K-selected catalogue by 0.2 dex. This difference originates at
luminosities lower than $L_\ast$ which implies that K-selected surveys
miss a significant fraction of star-forming lower-luminosity galaxies.\\
$\bullet$ At all redshifts, luminous galaxies ($L>L_\ast$) contribute
only $\sim \frac{1}{3}$ to the total SFR, i.e. the integrated SFR of
$L<L_\ast$ galaxies is a factor of $\sim 2$
higher than the one of $L>L_\ast$ galaxies.\\
$\bullet$ Our fits to the FDF luminosity functions suggest a flat
faint-end slope of $\alpha=-1.07 \pm 0.04$ in contrast to the assumed
slope of $\alpha \sim -1.6$ in the literature.  This implies that past
determinations have overestimated the SFR by a factor 2.\\
$\bullet$ The SFR is approximately constant over the redshift range
$1\le z\le4$ and drops by about 50\% around $z=4.5$, if dust
corrections constant with redshift are assumed.



\acknowledgments
We thank the anonymous referee for his helpful comments which
improved the presentation of the paper considerably.
This work was supported by SFB 375 of the DFG, by the German Ministry
for Science and Education (BMBF) through DESY under the project
05AE2PDA/8, and by the Deutsche Forschungsgemeinschaft under the
project SCHN 342/3--1 M. Schirmer und T. Erben thank C. Wolf for
providing some of the optical images used here.  Observations have
been carried out using the Very Large Telescope at the ESO Paranal
Observatory under Program ID(s): LP168.A-0485

\clearpage

\end{document}